# Direct experimental evidence of π magnetism of a single atomic vacancy in graphene

Yu Zhang[§], Si-Yu Li[§], Wen-Tian Li, Jia-Bin Qiao, Wen-Xiao Wang, Long-Jin Yin, and Lin He*

Center for Advanced Quantum Studies, Department of Physics, Beijing Normal University, Beijing, 100875, People's Republic of China

* Email: helin@bnu.edu.cn

**The pristine graphene is strongly diamagnetic. However, graphene with single carbon atom defects could exhibit paramagnetism with local magnetic moments ~ 1.5 $\mu_B$ per vacancy[1-6]. Theoretically, both the σ electrons and π electrons of graphene contribute to the magnetic moment of the defects, and the π magnetism is characterizing of two spin-split DOS (density-of-states) peaks close to the Dirac point[1,6]. Since its prediction, many experiments attempt to study this π magnetism in graphene, whereas, only a notable resonance peak has been observed around the atomic defects[6-9], leaving the π magnetism experimentally so elusive. Here, we report direct experimental evidence of the π magnetism by using scanning tunnelling microscope. We demonstrate that the localized state of the atomic defects is split into two DOS peaks with energy separations of several tens meV and the two spin-polarized states degenerate into a profound peak at positions with distance of ~ 1 nm away from the monovacancy. Strong magnetic fields further increase the energy separations of the two spin-polarized peaks and lead to a Zeeman-like splitting. The effective g-factors $g_{eff}$ around the atomic defect is measured to be about 40. Such a giant enhancement of the g-factor is attributed to the strong spin polarization of electron density and large electron-electron interactions near the atomic vacancy.**

According to Lieb's theorem, derived for a half-filled single-band Hubbard model, the ground state of materials with a bipartite lattice has magnetic moment $|N_A-N_B|\mu_B$, where $N_A$ and $N_B$ are the numbers of sublattice sites[10]. Graphene has the unique bipartite honeycomb lattice (the two sublattices are denoted by *A* and *B*). Therefore, a single carbon atom vacancy in graphene is expected to lead to a magnetic moment. First principles calculations predicted that an isolated vacancy in graphene with a planar configuration has a local magnetic moment of $1.5\mu_B$, in which $1\mu_B$ contributed by the dangling σ states and about $0.5\mu_B$ contributed by π electrons[1,11,12]. The π magnetism in graphene is closely linked to two spin-split DOS peaks near the Dirac point[1,11,12], which enables us to direct detect the signature of π magnetism on individual monovacancy by investigating its electronic structure. Very recently, the emergence of edge magnetism in individual zigzag graphene nanoribbon is also studied by measuring its electronic structure via scanning tunneling microscope (STM) [13-15].

To create atomic vacancies, the graphene systems are usually irradiated with high-energy ions[6-9] because of that the formation energy of a single atom vacancy in graphene is as high as about 7.4 eV[16,17]. Unfortunately, only a profound resonance peak has been observed on the irradiated vacancies[6-9]. The quenching of the magnetic moment of monovacancy in graphene is partially attributed to the graphene-substrate interaction[8]. Theoretically, a monovacancy in graphene with a metastable nonplanarity configuration is predicted to have a vanishing magnetic moment and the metastable configuration is only 50-100 meV above the ground state (i.e., the planar configuration)[8,12]. These irradiated vacancies generated by high-energy ions may not in the ground state. Therefore, the possible graphene-substrate interaction and the existence of the metastable configuration make the π magnetism of monovacancy in graphene quite elusive. In the current work, we directly synthesize graphene with high density of monovacancies on Rh foil using a facile ambient pressure chemical vapor deposition (CVD) method (more details are given in the Methods and Supplemental Fig. 1)[18-20]. We demonstrate here, using STM measurements, that the studied monovacancies are in the planar configuration. The spatial-resolved scanning tunneling spectroscopy (STS) spectra and the ground-state planar configuration of the monovacancies provide us unprecedented opportunities to direct measure the two spin-split DOS peaks of the π magnetism in graphene.

Figure 1a shows Raman spectroscopy mapping of a graphene multilayer on Rh foil. The regions with different colors reflect a change in the intensity of the Raman D peak (see supplemental Fig. 2), i.e., the density of defects in graphene[21]. The existence of high density of atomic-scale defects in graphene on Rh foil is further confirmed by STM measurements. Figure 1b shows a representative STM image of the studied graphene and tens of atomic-scale defects can be clearly identified. Such a high density of defects is attributed to a segregation mechanism for the graphene grown on Rh foil and the high carbon solubility of this metal[20]. In the STM measurements, the single carbon vacancies of graphene exhibit two distinct topographic signatures, as shown in Fig. 1c and Fig. 1d. For the single carbon vacancy on the topmost graphene layer, we observe the triangular $\sqrt{3}\times\sqrt{3}R30°$ interference pattern (Fig. 1c and supplemental Fig. 3), which is similar to that reported in previous studies[6-9]. Electronic contributions dominate the STM contrast near the monovacancy, making it difficult to resolve its atomic structure by STM. However, for the single carbon vacancy on the underlying graphene, the topmost graphene sheet reduces the contributions of electronic structure to the STM contrast and close-up topographic studies of such a defect could direct reveal its atomic structure, as shown in Fig. 1d. STM measurements at varying voltage bias show unblemished atomically resolved honeycomb structure of the topmost graphene layer above the single carbon vacancy (supplemental Fig. 4), which confirms that the defect in Fig. 1d is on the underlying graphene.

According to theoretical calculations, the single unpaired C atom of the monovacancy in graphene with the metastable non-magnetic configuration is about 50 pm out of the graphene plane[8,12]. To confirm that the studied single carbon vacancies are not in the non-magnetic configuration, we use different bias voltages to acquire the STM images of the defects on the topmost layer (supplemental Fig. 5). The height of the vacancy depends strongly on the bias, showing that the STM contrast originates primarily from the locally modified electronic structure rather than from topographic features. The structural origin of observed protrusions of the vacancies is estimated to be less than 20 pm. For graphene multilayer grown on Rh foil, there is usually rotational misalignment between the layers, resulting in moiré patterns in STM images (supplemental Fig. 6)[18,19,20]. In such a case, the topmost graphene sheet usually decouples from the underlying graphene systems and Landau quantization of massless Dirac fermions can be observed in the decoupled topmost graphene sheet[23-28], as demonstrated in Fig. 3. Therefore, we can also exclude any possible interaction between the single carbon vacancy and the substrate.

Figure 2a shows several representative spectra recorded around a single carbon vacancy in graphene on Rh foil. With approaching the monovacancy, a resonance peak emerges in the tunneling spectra, which is similar to that reported in previous studies[6-9]. A notable feature of the STS spectra is the splitting of the resonance peak within ~ 0.6 nm around the vacancy and its sensitive dependency on the recorded positions. Such a feature, which has never been reported before, reminds us of the characteristics of the π magnetism in graphene[1]. The tunneling spectrum gives direct access to the local DOS of the surface beneath the STM tip. The two peaks of the STS spectra are attributed to the DOS of the opposite spin polarizations and their amplitudes reflect the distribution of the electron spin density around the defect. Our result, as shown in Fig. 2a, indicates that the spin splitting is much more localized than the resonance peak of the single carbon vacancy. The two spin-polarized peaks merge into a single one at ~ 0.6 nm away from the monovacancy, implying that electron-electron interactions of the localized states play a vital role in the spin splitting. Similar observations are obtained in tens of single carbon vacancies in the graphene on Rh foils with many STM tips. Figure 2b shows several typical spectra of five different monovacancies in our studies. Consequently, our measurements verify the predicted two spin-polarized DOS peaks, *i.e.*, the π magnetism, of single carbon vacancy in graphene.

To further explore the electronic properties of the single carbon vacancies in graphene on Rh foils, we carried out STS measurements in various magnetic fields.

Figure 3a shows spectra recorded in high magnetic fields at position ~ 2.6 nm away from the defect. The spectra exhibit Landau quantization of massless Dirac fermions and the energies of the discrete Landau levels (LL) can be described by[23-27]

$$E_n = \text{sgn}(n)\sqrt{2e\hbar v_F^2 |n| B} + E_0, \quad n = ...-2,-1,0,1,2...\quad (1)$$

Here $E_0$ is the energy of Dirac point, $e$ is the electron charge, $n$ is the Landau index, $\hbar$ is the Planck's constant, and $v_F$ is the Fermi velocity. The observation of a sequence of LLs showing single-layer-graphene scaling demonstrates the efficient decoupling of the topmost graphene sheet from the supporting substrate. The linear fit of the experimental data to Eq. (1), as shown in Fig. 3b, yields the Fermi velocity of electrons $v_F^e = 1.18 \times 10^6$ m/s and the velocity of holes $v_F^h = 1.04 \times 10^6$ m/s. Such a large electron-hole asymmetry may arise from the local lattice deformation around the monovacancy and the electron scattering of the charged defect[29,30].

Besides the large electron-hole asymmetry, the other notable observation of the spectra is the lifting of the degeneracy of LLs in high magnetic fields, as shown in Fig. 3a. For example, the energy splitting of the $LL_0$ is about 5.5 meV in the field of 5 T and it increases to about 8.8 meV in the field of 7 T. Similar energy splitting is also observed in $LL_{-1}$ and $LL_1$ in the field of 7 T. The energy splitting of LLs with higher orbital indices is not observed in our experiment because of that the line width of LLs increases with energy. Such a behavior is related to the quasiparticle lifetimes, which decrease with the energy difference from the Fermi level[23,25]. Fitting the splitting energies to a Zeeman-like dependence, $E = g\mu_B B$, yields the g-factor $g \sim 21$. We attribute this energy splitting to the lifting of the valley degeneracies because the effective g-factor of the valley splitting in graphene is measured to be about 18.4[25]. In the presence of high magnetic field, the electrons are spatially more localized and the electron-electron interaction is expected to be enhanced. The enhanced interaction lifts LL degeneracies and generates gaps in graphene[25,31]. The observation of interaction-driven gaps, which increase with the magnetic fields, is a clear signature that the electron-electron interaction in graphene is enhanced with increasing the fields.

Figure 3b shows several representative spectra recorded in high magnetic fields at a monovacancy. We can detect weak signals of LLs at the monovacancy because of that the wavefunctions of LLs have their spatial extent, $\sim 2\sqrt{N}l_B$ (here $N$ is the Landau index and $l_B = \sqrt{\hbar/eB}$, which is of the order of 10 nm for the magnetic fields applied in our experiment). A notable result of the spectra is that the energy separations of the two spin-polarized DOS peaks of the monovacancy increase with the magnetic fields, as shown in Fig. 3b and summarized in Fig. 4. A linear fit the experimental data to $\Delta E = g^*\mu_{\text{eff}}B$ yields $g^* \sim 40$ (here we use $\mu_{\text{eff}} \sim 1.5\mu_B$). This effective g-factor is much larger than that of the valley splitting measured away from the defect, implying significant contribution of the monovacancy to the giant enhancement. Previously, a large effective g-factor $g \sim 34$ has been observed in electron puddles of graphene bilayer on $SiO_2$ substrate and such a large g-factor is attributed to many-body interactions[31]. In our case, the spin splitting of the two peaks of the monovacancy in graphene is determined by the strength of the electron-electron interaction. Thus, the enhanced interaction in high fields further increases the energy separations of the two spin-polarized peaks. We believe that the observed large *g*-factor in our experiment is due in part to the strong spin polarization of electron density of the monovacancy, and in part to the enhanced electron-electron interactions in high magnetic fields.

**Methods:**

A traditional ambient pressure chemical vapor deposition (APCVD) method was adopted to grow the atomic-scale defected graphene on Rh foil. The Rh foil was synthesized at 1000 °C at the atmosphere of methane ($CH_4$) and hydrogen ($H_2$) via the segregation mechanism. When the foil was cooled down, the dissolved carbon atoms segregated from bulk to surface and reconstructed the original honeycomb lattices. Atomic-scale defects, such as heptagon-pentagon topological defects, adatoms, atomic vacancies, and flower defects, are generated during the growth process.

The STM system was an ultrahigh vacuum scanning probe microscope (USM-1500S) from UNISOKU with the magnetic fields up to 8 T. All the STM and STS measurements were performed in the ultrahigh vacuum chamber (~$10^{-11}$ Torr) with constant-current scanning mode. The experiments were acquired at temperature ~4.5 K. The STM tips were obtained by chemical etching from a wire of Pt(80%) Ir(20%) alloys. Lateral dimensions observed in the STM images were calibrated using a standard graphene lattice and a Si (111)-(7×7) lattice and Ag (111) surface. The *dI/dV* measurements were taken with a standard lock-in technique by turning off the feedback circuit and using a 793-Hz 5mV a.c. modulation of the sample voltage.

**Acknowledgments**

This work was supported by the National Basic Research Program of China (Grants Nos. 2014CB920903, 2013CBA01603), the National Natural Science Foundation of China (Grant Nos. 11422430, 11374035), the program for New Century Excellent Talents in University of the Ministry of Education of China (Grant No. NCET-13-0054), Beijing Higher Education Young Elite Teacher Project (Grant No. YETP0238). L.H. also acknowledges support from the National Program for Support of Top-notch Young Professionals.

**Note added**: During the preparation of the manuscript, we became aware of the work of Gonzalez-Herrero, et al (32), which showed the existence of magnetism in graphene by using hydrogen atoms.

**Figure Captions:**

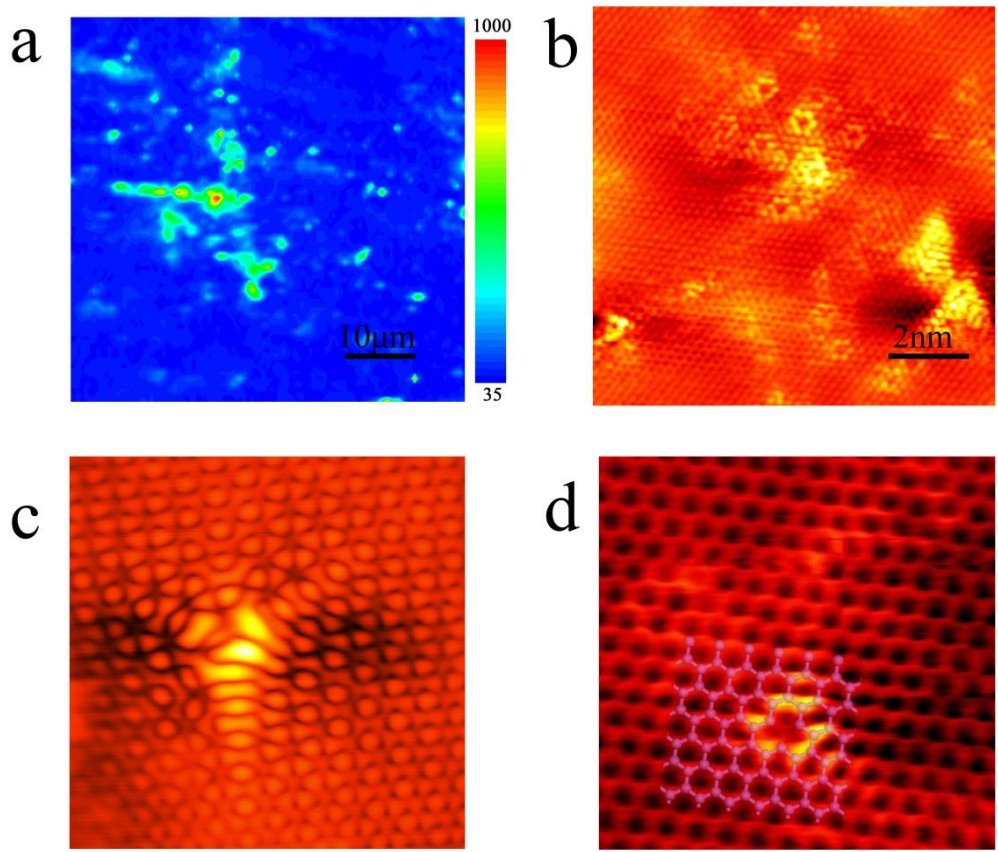

**Figure 1 | Graphene with atomic defects on Rh foil. a,** 46 μm×46 μm Raman mapping of D band (peak at 1350 cm$^{-1}$ of Raman spectrum) of graphene on Rh foil. **b,** 10 nm × 10 nm STM topographic image of graphene region with high density of defects on Rh foil. **c,** Atomic resolution STM image of a single carbon vacancy in topmost graphene sheet. **d,** Zoom-in STM images of a single carbon vacancy in the underlying graphene.

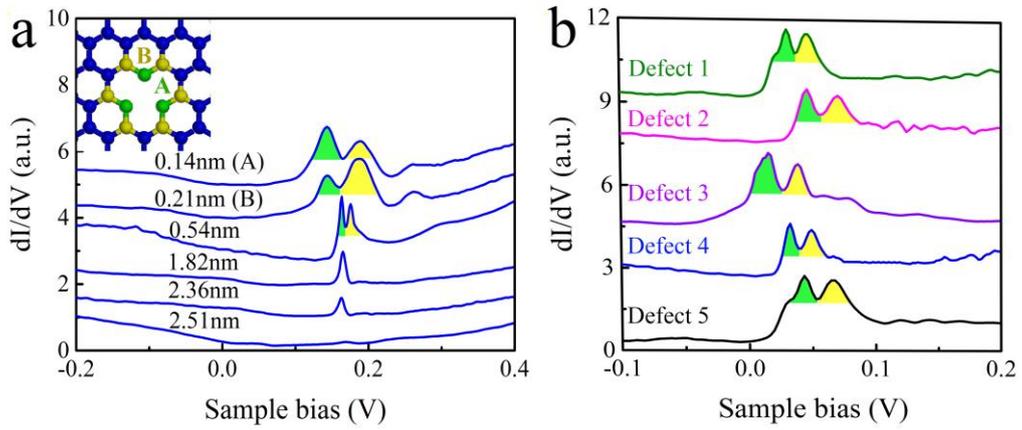

**Figure 2 | Spin-split states of single carbon vacancy in graphene. a,** STS spectra recorded at different distances away from a single carbon vacancy. The two peaks reflect the DOS with opposite spin polarizations. Inset: Schematic structure of a single carbon vacancy. The first and second spectra are measured on the nearest-neighbor and the next-nearest-neighbor of the monovacancy. **b,** Representative STS spectra of five different monovacancies. The energy separations of the two peaks $\Delta E$ vary from about 20 meV to about 60 meV.

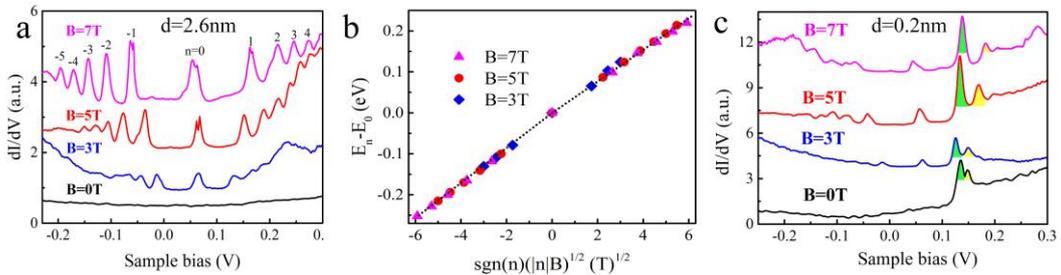

**Figure 3 | Landau quantization around the monovacancy. a,** Tunneling spectra recorded at position 2.6 nm away from a monovacancy under different magnetic fields. LL peak indices are labeled and the data are offset in the Y axis for clarity. A slight splitting of $LL_0$ and $LL_{\pm 1}$ can be observed in the field of 7 T. **b,** The energies of LLs show a linear dependence against $sgn(n)(|n|B)^{1/2}$, as expected for massless Dirac fermions in the graphene monolayer. The solid curves are linear fits of the data with Eq. (1). **c,** Tunneling spectra recorded at position ~ 0.2 nm away from the monovacancy under various magnetic fields. The energy separations of the two peaks

increase with the magnetic fields.

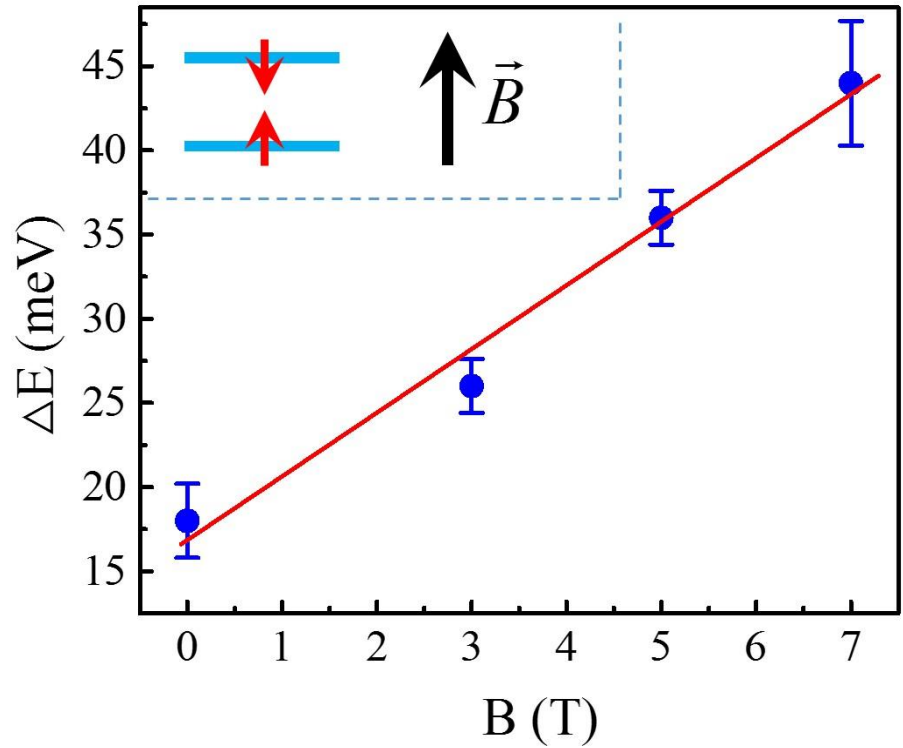

**Figure 4 | Energy separations of the two spin-split peaks of the monovacancy as a function of magnetic fields.** A linear fit of the energy separations versus magnetic field yields an effective *g*-factor of 40 around the monovacancy.